\documentstyle[12pt,epsfig,a4]{article} 
\newcommand{\vs}{\vspace{-0.25cm}}
\begin{document} 
\begin{center}
{\Large{\bf Particle-hole ring diagrams for fermions in two dimensions}}

\bigskip

N. Kaiser\,\footnote{ Email address: nkaiser@ph.tum.de}\\
\medskip
{\small Physik-Department T39, Technische Universit\"{a}t M\"{u}nchen,
   D-85747 Garching, Germany}
\end{center}
\medskip
\begin{abstract}
The set of particle-hole ring diagrams for a many-fermion system in two 
dimensions is studied. The complex-valued polarization function is 
derived in detail and shown to be expressible in terms of square-root functions. 
For a contact-interaction the perturbative contributions to the 
energy per particle $\bar E(k_f)$ are calculated in closed analytical form 
from third up to twelfth order. The resummation of the particle-hole ring 
diagrams to all orders is studied and a pronounced dependence on the 
dimensionless coupling parameter $\alpha$ is found. There is a substantial difference 
between the complete ring-sum with all exchange-type diagrams included and the standard 
resummation of the leading $n$-ring diagrams only. The spin factor $S_n(g)$ 
associated to the $n$-th order ring diagrams is derived for arbitrary 
spin-degeneracy $g$. 
\end{abstract}


\section{Introduction}
An important class of diagrams in finite-temperature and many-body 
calculations is the set of the particle-hole ring diagrams. In gauge theories 
the ring diagrams play a crucial role since they incorporate screening 
corrections to the (bare) infinite-range interactions and their inclusion 
becomes necessary in order to achieve a well behaved perturbative expansion 
at all \cite{lebellac}. In theories with short-range interactions the 
particle-hole bubbles provide also important corrections to the effective 
interaction. An example for this feature is the induced interaction in Landau's  
theory for normal Fermi-liquids \cite{baym,babu}. Likewise, the 
energies and lifetimes of collective excitations of a system (phonons, 
plasmons etc.) are determined from the poles of the particle-hole 
polarization propagator.  

The particle-hole ring diagrams have also been considered repeatedly in 
effective field theory approaches to (three-dimensional) many-fermion systems 
at finite density 
\cite{hammer,schaefer}. Based on a contact-interaction proportional to the  
s-wave scattering length $a$, the particle-hole ring diagrams contribute at third 
and higher orders to the systematic low-density expansion of the equation of state 
\cite{hammer2,steele}. The respective coefficients occurring in the expansion in 
powers of $ak_f$ are however known only with moderate numerical accuracy. 
Another feature of the particle-hole ring diagrams with a contact-interaction is 
that their spin-factor (a polynomial of degree $n$ in $g$) varies in a somewhat 
complicated way with the order $n$ and the spin-degeneracy $g$. For large 
spin-degeneracy $g$ the particle-hole ring diagrams are actually distinguished 
as the leading ones in a $1/g$-expansion (with $ag=$\,constant) and their 
resummation to all orders has been studied in detail in refs.\cite{hammer,schaefer}. 
Moreover, the resummed particle-hole ring diagrams could be used to derive 
in the Bose limit a non-perturbative result with a non-analytical dependence 
on the scattering length $a$ and the particle density \cite{hammer}. 

The purpose of the present paper is to study in the same way the particle-hole 
ring diagrams for fermions in two spatial dimensions. The analytical evaluation of 
fermionic in-medium ladder diagrams in two dimensions together with their resummation 
to all orders has been presented recently in (the appendix of) ref.\cite{2dlad}. 
The pertinent 
complex-valued polarization function $\Pi(q_0 ,|\bf q|)$ for fermions in two 
dimensions is rederived in section 2. One finds that both its real and imaginary 
part can be expressed in terms of simple square-root functions 
of two dimensionless variables. This represents an essential simplification in 
comparison to the three-dimensional case where logarithmic functions occur in the 
expression for the real part. Given the polarization function the perturbative 
contributions to the energy per particle $\bar E(k_f)$ as they arise from particle-hole 
ring diagrams with a contact-interaction are calculated in section 3. The dependence 
on the Fermi momentum is $k_f^2/M$ at each order and, most remarkably, the pertinent 
numerical coefficients can be given as exact numbers (involving rationals and $\ln 2$) 
from third up to twelfth order. The resummation of the two-dimensional particle-hole ring 
diagrams to all orders is then studied in section 4. One finds a dependence on the 
dimensionless coupling parameter $\alpha$, which is much more pronounced than the 
dependence on $a k_f$ in the three-dimensional case. Moreover, the complete ring-sum 
with all exchange-type diagrams included differs substantially from the resummation 
of the leading $n$-ring diagrams only. 
In the appendix a derivation of the formula $S_n(g)= (g-1)^n+(-1)^n(g^2-1)$ for 
the general spin-factor of $n$-th order particle-hole ring diagrams is presented. 

It is hoped that the results of the present paper will find applications in 
special condensed matter systems (such as two-dimensional graphene \cite{graphene} or 
strongly interacting  two-dimensional Fermi gases \cite{froehlich,randeria}).
 
\section{Polarization function}
\begin{figure}
\begin{center}
\includegraphics[scale=0.45,clip]{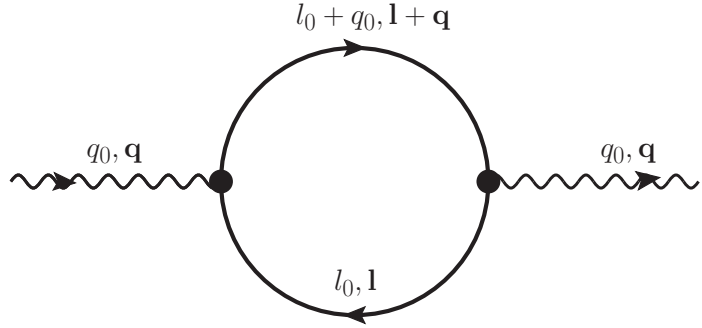}
\end{center}
\vspace{-.5cm}
\caption{Elementary particle-hole polarization bubble with energy $q_0$ and 
momentum $\bf q$ flowing through it. The wiggly line symbolizes the contact-interaction.}
\end{figure}

The basic quantity for treating the particle-hole ring diagrams is the complex-valued
polarization function $\Pi(q_0 ,|\bf q|)$, depending on energy $q_0$ and momentum 
$|\bf q|$. For non-relativistic fermions in three dimensions its explicit derivation 
can be found e.g. in the textbook \cite{fetter}. The expression for the polarizability of 
the two-dimensional electron gas has been given long ago by Stern in ref.\cite{stern}. 
Nevertheless, it is instructive to rederive the result for the two-dimensional 
polarization function by presenting some details of the calculation.   
In this work we consider spin-1/2 
fermions of mass $M$ in two dimensions and introduce a two-body contact-interaction 
with coupling constant $C_0= 2\pi \alpha/M$, where $\alpha$ is a dimensionless 
parameter. When including this coupling constant and a factor $-2$ from the sum 
over spins and the closed fermion loop in Fig.\,1, the  two-dimensional polarization 
function reads:
\begin{eqnarray} \Pi(q_0 ,|\bf q|) &\!\!\!=\!\!\!& -{4\pi i \alpha \over M}\! \int\!
{dl_0 \over 2\pi}\int\!{d^2{\bf l} \over (2\pi)^2} \bigg\{ {i \over l_0-{\bf l}^2
/2M +i \epsilon} -2\pi\, \delta(l_0-{\bf l}^2/2M)\, \theta(k_f-|\bf l|) 
\bigg\} \nonumber \\ && \times \bigg\{ {i \over l_0+q_0-({\bf l}+{\bf q})^2/2M 
+i \epsilon} -2\pi \,\delta(l_0+q_0-({\bf l}+{\bf q})^2/2M)\, \theta(k_f-
|\bf l+\bf q\,|) \bigg\}\,. \end{eqnarray}
Here, each factor in the curly brackets is a non-relativistic particle-hole propagator 
which has been rewritten identically as the sum of a vacuum-propagator and 
a medium-insertion. The Fermi momentum $k_f$ appearing in the step-functions 
$\theta(\dots)$ is related to the particle density (per area) by $\rho_2 = 
k_f^2/2\pi$.  By shifting the integration variable $l_0\to l_0-q_0/2$, one deduces 
immediately that $\Pi(q_0 ,|\bf q|)$ is an even function of the energy $q_0$. Therefore, 
one can assume  $q_0\geq 0$ in the following.  Next, the momentum-space integrals are 
evaluated for all four terms in eq.(1). The term with no step-function gives zero upon 
$l_0$-integration with residue calculus, 
since both poles lie in the lower $l_0$ half-plane. The two contributions with 
one medium-insertion lead to a real part of the form:
\begin{equation} {\rm Re}\, \Pi(q_0 ,|{\bf q}|) = 4\pi \alpha  -\hspace{-0.44cm}\int
{d^2{\bf l} \over (2\pi)^2}\bigg\{ { \theta(k_f-|{\bf l}|)\over {\bf l}\cdot {\bf q}
+ {\bf q}^2/2-Mq_0 } +(q_0\to -q_0)\bigg\}\,, \end{equation}
where we have substituted ${\bf l}\to -{\bf l}-{\bf q}$ in the second contribution. 
It is convenient to introduce dimensionless variables $(x, \nu)$ by setting $|{\bf q}| 
= 2x k_f$ and $q_0=2\nu x k_f^2/M$, where the extra factor $x$ in the parametrization of 
$q_0$ simplifies all following formulas. The remaining principal-value integral is 
readily solved by using planar polar coordinates:
\begin{eqnarray} && {\alpha \over 2\pi x}\int_0^1 d\lambda -\!\!\!\!\!\!\int_0^\pi 
d\varphi \,
\Big(x-\nu + \sqrt{\lambda} \cos\varphi\Big)^{-1} = {\alpha \over 2 x}\,{\rm sign}
(x-\nu)\,{\rm Re}\int_0^1 d\lambda \,\Big[(x-\nu)^2- \lambda \Big]^{-1/2} \nonumber \\ && =  
{\alpha \over x}\,{\rm sign}(x-\nu)\,{\rm Re}\Big\{|x-\nu|-\sqrt{(x-\nu)^2-1}\Big\}\,.
\end{eqnarray}
The second term in eq.(2) is included by adding the same expression with $\nu \to 
-\nu$. Scaling out the parameter $\alpha$, one gets finally for the real part of the 
two-dimensional polarization bubble $ {\rm Re}\,\Pi(q_0,|{\bf q}|)=\alpha\,R(\nu,x)$
with the function:
 \begin{equation} R(\nu,x) =2 -{1\over x} \,{\rm Re}\Big\{ {\rm sign}(x+\nu)
\sqrt{(x+\nu)^2-1}+  {\rm sign}(x-\nu)\sqrt{(x-\nu)^2-1}\,\Big\}\,. \end{equation}
The imaginary part of the two-dimensional polarization function has the representation:
\begin{eqnarray} {\rm Im}\, \Pi(q_0 ,|{\bf q}|) &\!\!\!\!\!\!\!=\!\!\!\!\!\!\!& \alpha  
\int\!d^2{\bf l} \, \delta(Mq_0-{\bf l}\cdot{\bf q})\Big\{\theta(k_f-|{\bf l}-{\bf q}/2|)
+ \theta(k_f-|{\bf l}+{\bf q}/2|) \nonumber \\ && \qquad \qquad  \qquad \qquad  \quad  
-2 \,\theta(k_f-| {\bf l}-{\bf q}/2|)\, \theta(k_f-|{\bf l}+{\bf q}/2|)\Big\}\,,
\end{eqnarray} 
where the shift $ {\bf l}\to {\bf l}- {\bf q}/2$ has been made.
By analyzing for the first term in eq.(5) the intersection of a straight line (as imposed 
by $\delta(Mq_0-{\bf l}\cdot{\bf q})$) and a shifted circular disc in the $\bf l$-plane, 
one finds that it contributes to the imaginary part with the term: $\alpha\, x^{-1} 
\sqrt{1-(x-\nu)^2}$ for $x-1<\nu<x+1$. Via the substitution $\bf l \to -\bf l$ 
the contribution of the second term in eq.(5) is obtained by altering the sign of $\nu$ 
and thus it becomes: $\alpha\, x^{-1} \sqrt{1-(x+\nu)^2}$ for $-x-1<\nu<1-x$. Since 
only positive $\nu$ need to be considered the latter condition is equivalent to 
$0<\nu<1-x$. In order that the third term in eq.(5) contributes at all, the two 
shifted Fermi discs must overlap, which requires $0<x<1$. In this case the third term 
subtracts twice the second contribution in the region  $0<\nu<1-x$, and thus 
effectively reverses the sign of that second contribution. Putting all three pieces 
together and scaling out $\alpha$ again, one obtains for the imaginary part of the 
two-dimensional polarization bubble 
${\rm Im}\, \Pi(q_0,|{\bf q}|) = \alpha \, I(\nu,x)$ with the function:
\begin{eqnarray}&& I(\nu,x) = {1\over x} \sqrt{1-(x-\nu)^2}\,, \quad  {\rm for} 
\quad |x-1|<\nu <x+1\,, \nonumber \\ && I(\nu,x) = {1\over x}\bigg\{ \sqrt{1-
(x-\nu)^2}-\sqrt{1-(x+\nu)^2}\bigg\}\,, \quad  {\rm for} \quad 0<\nu <1-x\,. 
\end{eqnarray}
It is interesting to note that the imaginary part of three-dimensional polarization 
function is of similar structure \cite{hammer,schaefer,fetter}. The parameter $\alpha$ 
gets replaced by $a k_f/2$ and the individual terms in eq.(6) come without square-roots. 
The real part of the three-dimensional polarization function is composed of logarithms: 
$\ln[(1+x+\nu) / |1-x-\nu|]$ and $\ln[|1+x-\nu| / |1-x+\nu|]$. One should also note that 
the Lindhard functions of a Fermi gas in one, two, and three dimensions have been studied 
extensively and compared with each other in ref.\cite{lindhard}. After translating the 
dimensionless variables $(\nu,x)$ the present results for $R(\nu,x)$ and $I(\nu,x)$ 
written in eqs.(4,6) agree exactly with the earlier calculations in 
refs.\cite{stern,lindhard}.

Fig.\,2 shows the support of the function $I(\nu,x)$ in the (positive) $x\nu$-plane. 
The corresponding region lies in between the two straight lines $\nu = x-1$ and 
$\nu = x+1$. When crossing the (dashed) line $\nu = 1-x$, the function $I(\nu,x)$
changes its form and it vanishes along the circumference (with the exception of the 
line-segment $x=0,\, 0<\nu<1$). We anticipate that 
the real-valued integrand relevant for calculating the interaction energy density from 
particle-hole 
ring diagrams involves a proportionality factor $I(\nu,x)$. Exploiting this property, 
the function $R(\nu,x)$ can be restricted to the support of $I(\nu,x)$. Inspection 
of eq.(4) shows that the last term can be dropped, since its radicand 
$[\nu-(x+1)][\nu-(x-1)] $ is negative. Furthermore, in the region $0<\nu <1-x$ 
the other radicand $[\nu+(x+1)][\nu-(1-x)]$ becomes also negative. Altogether, the 
function $R(\nu,x)$ simplifies considerably to:  
\begin{eqnarray}&& R(\nu,x) = 2-{1\over x} \sqrt{(x+\nu)^2-1}\,, \quad  {\rm for} 
\quad |x-1|<\nu <x+1\,, \nonumber \\ && R(\nu,x) = 2 \,, \quad  {\rm for} \quad 
0<\nu <1-x\,,\quad 0<x<1\,, \end{eqnarray}
inside the relevant integration region.
\begin{figure}
\begin{center}
\includegraphics[scale=0.8,clip]{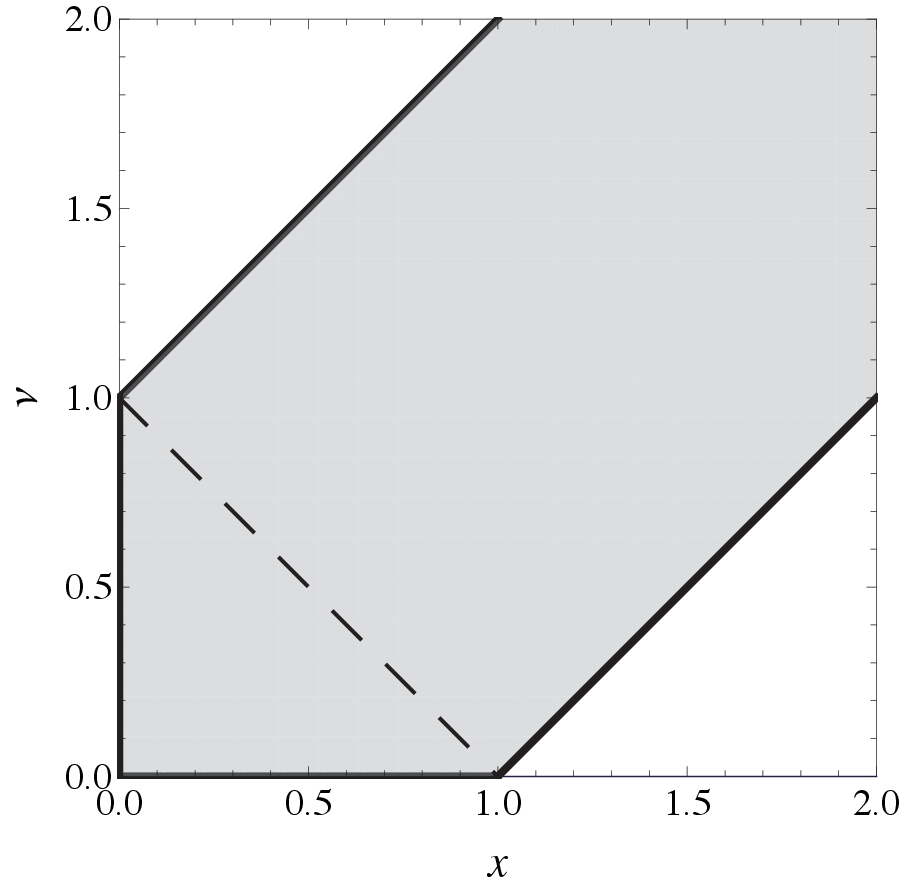}
\end{center}
\vspace{-.5cm}
\caption{Integration region in the $x\nu$ quarter-plane in between the two straight
lines $\nu=x-1$ and $\nu = x+1$. The functions $R(\nu,x)$ and $I(\nu,x)$ 
have different forms below and above the dashed line $\nu=1-x$.} 
\end{figure}

\section{Energy per particle and perturbative results}
In this section we calculate the energy per particle  $\bar E(k_f)$ as it arises from 
ring diagrams with a contact-interaction. For orientation we note that the free Fermi 
gas energy in two dimensions is $\bar E(k_f)^{(0)}= k_f^2/4M$. Likewise, the first-order 
Hartree-Fock contribution from a contact-coupling $C_0=2\pi \alpha/M$ is given by $\bar 
E(k_f)^{(1)}= -\alpha\,k_f^2/4M$. Obviously, the energy density arising from the leading 
$n$-ring diagram (obtained by connecting $n$ polarization bubbles in Fig.\,1 to a closed 
ring) is determined by the $n$-th power of the polarization function 
$\Pi(q_0,|{\bf q}|)$. With inclusion of the symmetry factor $1/2n$ the corresponding 
energy density reads: 
\begin{equation} {k_f^2 \over 2\pi} \, \bar E(k_f)^{n-{\rm ring}}_{\rm lead}  = {i \over 2n} 
\int\!{dq_0 \over 2\pi}\int\!{d^2{\bf q} \over (2\pi)^2} \,\Big[ \Pi(q_0,|{\bf q}|)\Big]^n
\,. \end{equation}
After transforming to the dimensionless variables $(x, \nu)$ one obtains the following 
double-integral representation for the contribution of $n$-ring diagrams to the energy 
per particle: 
\begin{equation}\bar E(k_f)^{n-{\rm ring}} = {4k_f^2 \over \pi M} {\alpha^n \over n}{S_n 
\over 2^n} \int_0^\infty \!\!dx\, x^2 \int_{\nu_{\rm min}}^{x+1}\!\! d\nu\, {\rm Im} 
\Big[R(\nu,x)-i I(\nu,x)\Big]^n  \,, \end{equation}
with $\nu_{\rm min} = {\rm max}(0,x-1)$. In addition to the leading $n$-ring diagram 
with spin-factor $2^n$, there are various exchange-type ring-diagrams due to the 
Pauli exclusion-principle. For a contact-interaction these can all be summed together 
by means of the total spin-factor $S_n = 1+3(-1)^n$. The corresponding sequence is 
$S_n=(-2,4,-2,4,\dots)$ \cite{schaefer}. In the appendix an explicit derivation of 
the formula $S_n = 1+3(-1)^n$, using multiple spin-traces, is given. The generalization 
to arbitrary spin-degeneracy $g\geq 2$ is also presented there.

Based on the representation in eq.(9) the respective double-integrals have been studied 
in detail by analytical and numerical methods. From third up to twelfth order analytical 
results in closed form could be obtained for the contributions to the energy per particle 
$\bar E(k_f)$. We list them one after the other:   

\noindent
$n=3$: the integrand $I(I^2-3R^2)$ leads to the 3-ring contribution:
\begin{equation}\bar E(k_f)^{3-{\rm ring}}={k_f^2\over 4 M}\alpha^3(2\ln2-1)\,,\end{equation}
$n=4$: the integrand $4 IR(I^2-R^2)$ leads to the 4-ring contribution:
\begin{equation}\bar E(k_f)^{4-{\rm ring}}={k_f^2\over M}\alpha^4(2-3\ln2)\,,\end{equation}
$n=5$: the integrand $I(10I^2R^2-5R^4-I^4)$ leads to the 5-ring contribution:
\begin{equation}\bar E(k_f)^{5-{\rm ring}}={k_f^2\over M}\alpha^5\bigg(4\ln2-{11\over 4}
\bigg)\,,\end{equation}
$n=6$: the integrand $2IR(3R^2-I^2)(3I^2-R^2)$ leads to the 6-ring contribution:
\begin{equation}\bar E(k_f)^{6-{\rm ring}}={k_f^2\over M}\alpha^6\bigg({83\over 6}-20\ln2
\bigg)\,,\end{equation}
$n=7$: the integrand $I^7+7I R^2(5I^2R^2-3I^4-R^4)$ leads to the 7-ring contribution:
\begin{equation}\bar E(k_f)^{7-{\rm ring}}={k_f^2\over M}\alpha^7\bigg(24\ln2-{133\over 8}
\bigg)\,,\end{equation}
$n=8$: the integrand $8IR(I^2-R^2)(I^4-6 I^2R^2+R^4)$ leads to the 8-ring contribution:
\begin{equation}\bar E(k_f)^{8-{\rm ring}}={k_f^2\over M}\alpha^8\bigg({4657\over 60}-112\ln2
\bigg)\,,\end{equation}
$n=9$: the integrand $I(I^2-3R^2)(33I^4R^2-I^6-27I^2R^4+3R^6)$ leads to the 9-ring 
contribution:
\begin{equation}\bar E(k_f)^{9-{\rm ring}}={k_f^2\over M}\alpha^9\bigg(128\ln2-{5323\over 60}
\bigg)\,,\end{equation}
$n=10$: the integrand $2IR(10I^2R^2-5I^4-R^4)(I^4-10I^2R^2+5R^4)$ leads to the 
10-ring contribution:
\begin{equation}\bar E(k_f)^{10-{\rm ring}}={k_f^2\over M}\alpha^{10}\bigg(
{27947\over 70}-576\ln2 \bigg)\,,\end{equation}
$n=11$: the integrand $I^{11}+11I R^2(30I^6R^2-5I^8-42I^4R^4+15I^2R^6-R^8)$ leads to 
the 11-ring contribution:
\begin{equation}\bar E(k_f)^{11-{\rm ring}}={k_f^2\over M}\alpha^{11}\bigg(640\ln2
-{149053\over 336}\bigg)\,,\end{equation}
$n=12$: the integrand $4I R(I^2-R^2)(I^2-3R^2)(3I^2-R^2)(I^4-14I^2R^2+R^4)$ leads to 
the 12-ring contribution:
\begin{equation}\bar E(k_f)^{12-{\rm ring}}={k_f^2\over M}\alpha^{12}\bigg({4918777\over 2520}
-2816\ln2\bigg)\,.\end{equation}
Following the pattern of numerical coefficients in eqs.(10-19) one conjectures for the
$n$-ring contribution:
\begin{equation}\bar E(k_f)^{n-{\rm ring}}={k_f^2\over M}(-\alpha)^n\Big\{\zeta_n -(n-1)2^{n-6}
[3+(-1)^n]\ln2\Big\}\,,\end{equation}
with $\zeta_n $ a positive rational number. In the course of the calculation the 
following observations have been made. The contribution to the double-integral in eq.(9) 
from the (triangular) region $0<\nu<1-x$ (see Fig.\,2) can be solved analytically, 
with a result equal to $\pi$ times a rational linear combination of $1$ and $\ln 2$. 
The integral over the (rectangular) region $|x-1|<\nu<x+1$ is numerically extremely close 
to a rational multiple of $\pi$. The corresponding values in the case $n=3$ are $\pi(7-12
\ln 2)/8$ and $-\pi/8$. In the cases $n = 4, 5, 6, 7$ the numerical integral over the 
rectangular region $|x-1|<\nu<x+1$ vanishes with an accuracy $10^{-10}$ or better,\footnote{
For the analytical and numerical treatment of the integrals is it occasionally useful to 
introduce new coordinates by $x = (\xi+\eta)/2$,  $\nu = (\xi-\eta)/2$. The integration 
region in the $\xi\eta$-plane consists then of a triangle $-\xi<\eta<\xi$, $0<\xi<1$ and a 
semi-infinite strip $-1<\eta<1$, $\xi>1$.} while for $n=8$ it has the value $-256\pi/15$. 
For $n\geq 9$ the integrals over both regions develop divergences which cancel 
in the sum. It is then advantageous to map the entire integration region onto a unit square 
$0<x,y<1$ via substitutions: $\nu = (1-x)y$, $\nu=1-x+2x y$, and $\nu=x-1+2 y,\, 
x \to1/x$. In order to illustrate the available precision we note that the large 
integers $4657,\,  5323,\, 27947,\,149053$ and $4918777$ in eqs.(15-19) are 
reproduced with a relative accuracy of $10^{-13}$ and better. The list of numerical coefficients 
$c_n$ for the $n$-ring contribution to the energy per particle 
 $\bar E(k_f)^{n-{\rm ring}} = c_n \alpha^n k_f^2/M$ is continued up to $n=30$ 
in Table\,1. One observes that the sequences of positive and 
negative coefficients $c_n$ decrease both slowly in magnitude.   

\begin{table}
\begin{center}
\begin{tabular}{|c|c c c c c c c|}\hline
$n$ & 3 & 4 & 5 & 6 & 7 & 8 &9 \\ \hline 
$100\, c_n$ &  $9.657359$  & $-7.944154$ & $2.258872$ & 
$-2.961028$ & $1.053233$ & $-1.581756$ & $0.617245$\\ \hline 
$n$ & 10  & 11& 12 &13 & 14 & 15& 16\\ \hline 
$100 \, c_n$ & $-0.991886$ & $0.407651$ & $-0.682554$ & 
$0.290058$ & $-0.499344$ & $0.217243$ & $-0.381589$         \\ \hline 
$n$ & 17  & 18& 19 &20 & 21 & 22& 23\\ \hline 
$100 \, c_n$ & $0.168933$ & $-0.301301$ & $0.135201 $ & 
$-0.244050 $ & $0.110695$ & $-0.201763 $ & $0.092323$         \\ \hline 
$n$ & 24  & 25& 26 &27 & 28 & 29& 30\\ \hline 
$100 \, c_n$ & $-0.169628$ & $0.078189$ & $-0.144627 $ & 
$0.067079 $ & $-0.124789$ & $0.058186 $ & $-0.108781$         \\ \hline 
\end{tabular}
\caption{Numerical coefficients $c_n$ of the energy per particle 
$\bar E(k_f)^{n-{\rm ring}} = c_n \alpha^n k_f^2/M$ arising from the $n$-th 
order particle-hole ring diagrams with a contact-interaction $2\pi \alpha/M$.}
\end{center}
\end{table}

Let us also give an alternative representation for the ring energy 
$\bar E(k_f)^{n-{\rm ring}}$ at $n$-th order. When following the method described 
in the text book by Gross, Runge and Heinonen \cite{grossrunge} and evaluating the 
key quantity defined in eq.(22.15) in two dimensions, one obtains:
\begin{equation} \bar E(k_f)^{n-{\rm ring}}= -{4k_f^2 \over \pi M}
{S_n\alpha^n \over n} \int_0^\infty\!\! dx\, x^2 \int_0^\infty\!\!dy\, 
[K(x,y)\big]^n \,, \end{equation} 
with the (two-dimensional) euclidean polarization function:
\begin{equation} K(x,y) = 1-{1\over \sqrt{2} x} \sqrt{x^2-y^2-1+
\sqrt{(1+x^2+y^2)^2-4x^2}} = 1 - {1\over x }\Big|{\rm Re} 
\sqrt{(x+i y)^2-1} \Big|\,.\end{equation} 
Note that the absolute magnitude in the short form removes the sign-ambiguity 
of the complex square-root. The function $K(x,y)$ takes on values in the range $0<K(x,y)<1$
and its asymptotic behavior in the $xy$ quarter plane is:
\begin{equation} K(r \cos\varphi, r \sin \varphi) = {1\over 2r^2} 
+{1-4 \sin^2 \varphi \over 8 r^4} +{\cal O}(r^{-6})\,.  \end{equation} 
Using the representation in eq.(21) all previous results 
collected in eqs.(10-19) could be reproduced with very high numerical accuracy.
\section{Resummation to all orders}
After the perturbative analysis of the two-dimensional ring diagrams with a 
contact-interaction we study now their resummation to all orders. Returning to eqs.(8,9) 
one sees that the resummation to all orders is provided by an arctangent-function 
with two subtractions. The subtraction terms (linear and quadratic in $\alpha$) introduce 
divergences (quadratic and logarithmic in the upper limit of the $x$-integration) which 
get canceled by the integral over the arctangent-function. In order to avoid such 
numerically delicate cancellations one better represents the derivative of $\bar E(k_f)$ 
with respect to $\alpha$ in terms of a convergent integral over a rational expression and 
integrates back in $\alpha$:   
\begin{eqnarray} \bar E(k_f)_{\rm lead} &\!\!\!\!\!\!\!=\!\!\!\!\!\!\!& {4k_f^2 \over \pi M} 
\int_0^\alpha\!\! d z\, z^2\int_0^\infty \!\!dx\, x^2 \int_{\nu_{\rm min}}^{x+1}\!\! 
d\nu\, I(\nu,x)\nonumber \\ && \times { I(\nu,x)^2-3 R(\nu,x)^2 + 2z R(\nu,x)
[ R(\nu,x)^2+I(\nu,x)^2] \over [1-z  R(\nu,x)]^2 +z^2   I(\nu,x)^2} = 
{k_f^2  \alpha^3\over 4M}\,  G(\alpha) \,.  \end{eqnarray}
Here, we are considering first the sum of the leading $n$-ring diagrams with 
spin-factors $2^n$. The dimensionless function $G(\alpha)$ is defined by scaling out 
the free Fermi gas energy $k_f^2/4M$ and the leading power $\alpha^3$. Its non-vanishing 
value at $\alpha=0$ is: $ G(0) = 4-8\ln 2=-1.54518$.

\begin{figure}[h]
\begin{center}
\includegraphics[scale=0.45,clip]{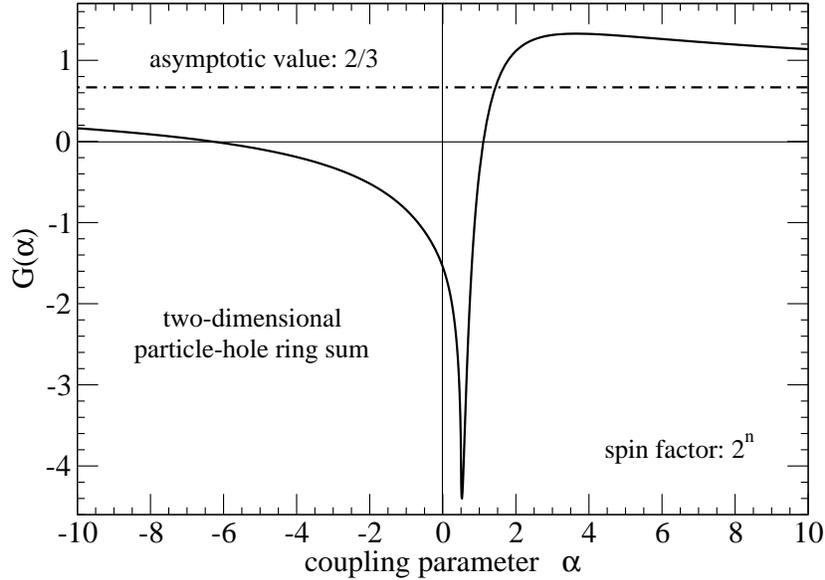}
\end{center}
\vspace{-.5cm}
\caption{The function $G(\alpha)$ resulting from the resummed particle-hole 
ring diagrams.}
\end{figure}

Fig.\,3 shows the behavior of the function $G(\alpha)$ in the range 
$-10< \alpha<10$ of the coupling parameter $\alpha$. One observes a sharp spike in 
the vicinity of $\alpha=1/2$. The function $G(\alpha)$ crosses zero twice, at 
$\alpha \simeq -6.33$ and  $\alpha \simeq 1.11$, and in the region $|\alpha|>2$ its 
variation becomes rather weak. A detailed numerical study at large positive and 
negative parameter values reveals that the asymptotic behavior of $G(\alpha)$ is given by: 
\begin{equation}  G(\alpha) \simeq {2 \over 3} +{2\ln|4\alpha|-5/2\over \alpha} 
\,. \end{equation}
This functional form has been found through a fit to the numerical data at large 
$|\alpha|$. The behavior of the function $G(\alpha)$ in the region $0<\alpha<1$ around 
the spike is shown with 
magnification in Fig.\,4. One makes the interesting numerical observation that $G(1/4) 
= -2$ and $G(1/2) = -4$ with an accuracy of $10^{-9}$. Furthermore, one sees that 
$G(\alpha)$ decreases further and reaches its minimum value of about $-4.40$ at the 
parameter value $\alpha \simeq 0.523$. One also recognizes from the magnified 
representation in Fig.\,4 that the behavior of $G(\alpha)$ around the ``spike'' 
is actually completely smooth. All these observed properties together with the results in 
eqs.(10-19) lead to the suggestion that the function $G(\alpha)$ could possibly be 
expressed through some decent analytical formula (involving presumably elliptic or 
hypergeometric functions). We have however not succeeded in finding such an analytical 
formula.

\begin{figure}[h]
\begin{center}
\includegraphics[scale=0.45,clip]{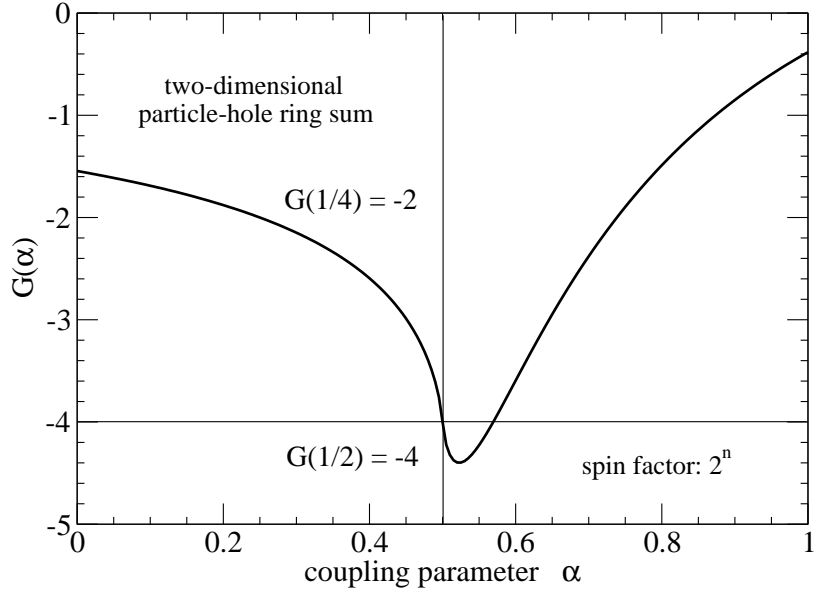}
\end{center}
\vspace{-.5cm}
\caption{The function $G(\alpha)$ magnified in the region $0<\alpha<1$. }
\end{figure}

Next, we consider the complete ring-sum with inclusion of all exchange-type diagrams.  
Based on the spin-factor $S_n= 1+3(-1)^n$ at $n$-th order (see eq.(9)) one obtains the 
following representation of the resummed energy per particle:
\begin{eqnarray} \bar E(k_f) &\!\!\!\!\!\!\!=\!\!\!\!\!\!\!& {4k_f^2 \over \pi M} 
\int_0^\infty \!\!dx\, x^2 \int_{\nu_{\rm min}}^{x+1}\!\! 
d\nu\, \bigg\{\alpha I(\nu,x)\big[\alpha R(\nu,x)-1\big] \nonumber \\ && + \arctan
{\alpha I(\nu,x)\over \alpha R(\nu,x)-2} +3 \arctan{\alpha I(\nu,x) \over \alpha 
R(\nu,x)+2} \bigg\} = {k_f^2 \alpha^3 \over 4M}\,  F(\alpha) \,.  \end{eqnarray}
The function $F(\alpha)$ defined by scaling out the free Fermi gas energy $k_f^2/4M$ 
and the leading power $\alpha^3$ is not independent. The relation to the function 
$G(\alpha)$ introduced before in eq.(24) reads: 
\begin{equation} F(\alpha) = {1\over 8}\Big\{G(\alpha/2)-3\,G(-\alpha/2)\Big\}\,. 
\end{equation} 
Fig.\,5  shows the behavior of the function $F(\alpha)$ in the range $-10< 
\alpha<10$ of the coupling parameter $\alpha$. One observes now two peaks located 
at $\alpha\simeq \pm 1 $, which simply arise from the minima of $G(\pm \alpha/2)$ 
(see Figs.\,3,4). The value at $\alpha=0$ is $F(0)=2\ln2- 1=0.386294$ and 
asymptotically $F(\pm \infty)=-1/6$ is reached. In order to reveal the strong parameter 
dependence in the region $-4< \alpha<4$ it is instrumental to scale out the 
trivial factor $\alpha^3$ in eq.(26). When scanning through from large 
negative to large positive values of coupling constant $\alpha$, the energy per 
particle $\bar E(k_f)$ changes its sign four times. This is a rather striking parameter 
dependence. Altogether, the main finding of the present work is that  complete 
ring-sum with all exchange-type diagrams included leads to a quantitatively and  
qualitatively different result for the energy per particle $\bar E(k_f)$ in comparison 
to the standard resummation of the leading $n$-ring diagrams.  

\begin{figure}[h!]
\begin{center}
\includegraphics[scale=0.45,clip]{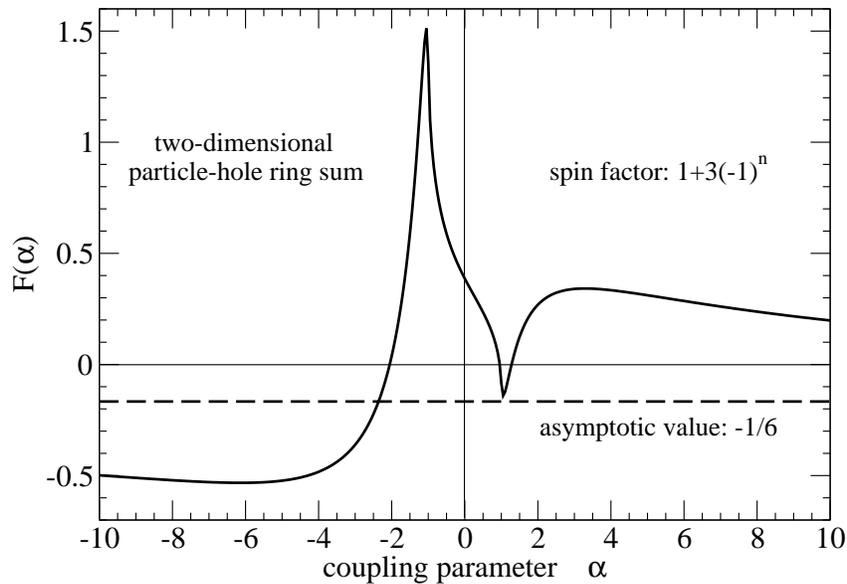}
\end{center}
\vspace{-.5cm}
\caption{The function $F(\alpha)$ resulting from the resummed particle-hole 
ring diagrams.}
\end{figure}

\begin{figure}[h]
\begin{center}
\includegraphics[scale=0.45,clip]{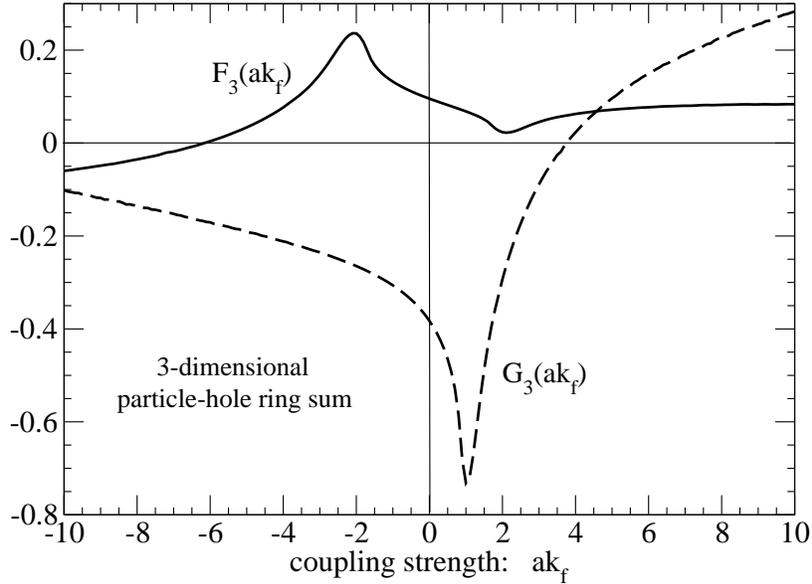}
\end{center}
\vspace{-.5cm}
\caption{The functions $G_3(a k_f)$ and $F_3(a k_f)$ resulting from the resummed 
particle-hole ring diagrams with a contact-interaction $4\pi a/M$ in three dimensions.}
\end{figure}

It is also interesting to compare with the analogous results for the resummed ring 
diagrams in three dimensions \cite{hammer,schaefer}. Using the product $a k_f$ of 
scattering length $a$ and Fermi momentum $k_f$ as a dimensionless coupling 
strength\footnote{The sign-convention is chosen such that $a>0$ corresponds to 
attraction.} and appropriate polarization functions $I_3(\nu,x)$ and $R_3(\nu,x)$, 
the only difference to eqs.(24,26) is an additional factor $3x$. The functions 
$G_3(ak_f)$ and $F_3(ak_f)=[G_3(ak_f/2)-3G_3(-ak_f/2)]/8$ are defined 
in the same way by dividing the resummed energy per particle $\bar E(k_f)$ by 
the three-dimensional free Fermi gas energy $3k_f^2/10M$ and the leading power 
$(ak_f)^3$. These two functions are shown by the dashed and full line in Fig.\,6 for
$-10<ak_f<10$. One observes a behavior similar to the one in
Figs.\,3,5, however with a less pronounced dependence on the dimensionless 
coupling strength $ak_f$. The 
function $G_3(ak_f)$ reaches its minimum of $-0.732$ at $a k_f \simeq 1.0$ and it 
crosses zero at  $a k_f \simeq 3.75$. The 
asymptotic behavior of $G_3(ak_f)$ is not obvious from Fig.\,6 and also very difficult 
to extract by numerical methods \cite{hammer}. One can conclude that the two-dimensional 
particle-hole ring diagrams with their multitude of analytical results represent a rather  
special case.   
\section{Summary and conclusions}
In this work we have studied the particle-hole ring-diagrams for a many-fermion in two
dimensions. Based on a contact-interaction the perturbative contributions to the energy 
per particle $\bar E(k_f)$ could be given in closed analytical from third up to twelfth 
order. The resummation to all orders exhibits a pronounced dependence on the 
dimensional coupling parameter $\alpha$. A particular finding of the present work has been
that the complete ring-sum with all exchange-type diagrams included (corresponding to 
spin-factors $1+3(-1)^n$) leads to a significantly different result for the energy per 
particle than the standard resummation of the leading $n$-ring diagrams 
(with spin-factors $2^n$). 

\section*{Appendix: Calculation of general spin-factor}
In this appendix we present a derivation of the general spin-factor associated to
the particle-hole ring diagrams at $n$-th order. We start with the simplest case of 
spin-1/2 fermions, i.e. spin-degeneracy $g=2$. While the leading $n$-ring diagram has 
the obvious spin-factor $g^n = 2^n$, the problem consists in determining this factor 
for all the exchange-type diagrams. For a four-fermion contact-interaction the 
exchange-type diagrams required by the Pauli exclusion-principle are automatically 
taken into account if the antisymmetrized form of the contact-interaction is used:
\begin{equation} {\bf 1} - {1\over 2}( {\bf 1}+\vec \sigma_1\cdot \vec \sigma_2) = 
{1\over 2}( {\bf 1}-\vec \sigma_1\cdot \vec \sigma_2)\,, \end{equation}
as obtained by subtracting the spin-exchange operator. The resulting operator 
$( {\bf 1}-\vec \sigma_1\cdot \vec \sigma_2)/2$ acts in the product spin-space of particle $1$ 
and particle $2$. The total spin-factor $S_n$ of the $n$-ring diagrams is now given 
by the trace of the product of $n$ such operators multiplied in cyclic order of their 
particle indices. The starting expression reads:
\begin{eqnarray} S_n &\!\!\!=\!\!\!& {1\over 2^n} {\rm tr}_1 {\rm tr}_2
\dots  {\rm tr}_n\{( {\bf 1}-\vec \sigma_n\cdot \vec \sigma_1)( {\bf 1}-\vec \sigma_1\cdot 
\vec \sigma_2)\dots ( {\bf 1}-\vec \sigma_{n-1}\cdot \vec \sigma_n)\}\nonumber \\ & \!\!\!=
\!\!\!&  {2\over 2^n} {\rm tr}_2\dots  {\rm tr}_n\{( {\bf 1}+\vec \sigma_n\cdot 
\vec \sigma_2)\dots ( {\bf 1}-\vec \sigma_{n-1}\cdot \vec \sigma_n)\}\,, \end{eqnarray}
and we have performed in the second line the trace in the spin-space of particle $1$. 
Note the sign change in front of the spin-spin operator $\vec \sigma_n\cdot 
\vec \sigma_2$. After taking $n-1$ spin-traces one arrives at: 
\begin{equation}S_n=  
{2^{n-1}\over 2^n} {\rm tr}_n\{ {\bf 1}+(-1)^n\vec \sigma_n\cdot \vec \sigma_n\} = 
1+3 (-1)^n\,, \end{equation}
where the factor $3$ stems from $\vec \sigma^{\,2}= {\bf 3} $. The same sequence of 
numbers $S_n=(-2,4,-2,4,\dots)$ has been reported in section IV of ref.\cite{schaefer} 
without a derivation. 

Next, we consider fermions with $g$ internal degrees of freedom. These ``flavor'' degrees 
of freedom are assumed to form the $g$-dimensional representation of the 
special unitary group $SU(g)$. The antisymmetrized contact-interaction which incorporates 
the Pauli exclusion-principle is now obtained by subtracting from the unit-operator 
${\bf 1}$ in two-particle flavor-space its Fierz transform:
\begin{equation} {\bf 1} - \bigg({{\bf 1}\over g}+{1\over 2}\vec \lambda_1\cdot 
\vec \lambda_2 \bigg)\,. \end{equation}
Here, $\vec \lambda$ denotes $g^2-1$ traceless generators that are normalized as 
tr$(\lambda^a\lambda^b) = 2 \delta^{ab}$, with indices $a,b \in\{1,\dots, g^2-1\}$. The 
operator ${\bf 1}/g+\vec \lambda_1\cdot \vec \lambda_2/2$ has indeed the properties of a 
flavor-exchange operator. Its eigenvalue in the $g(g+1)/2$-dimensional representation of 
symmetric tensors is $1/g+ (g+2)(g-1)/g- (g^2-1)/g =1$, and its eigenvalue  in the 
$g(g-1)/2$-dimensional representation of antisymmetric tensors is $1/g+ (g-2)(g+1)/g- 
(g^2-1)/g =-1$. The starting expression for the spin-factor $S_n(g)$ reads: 
\begin{eqnarray} S_n(g) &\!\!\!=\!\!\!& {\rm tr}_1 {\rm tr}_2\dots  {\rm tr}_n\bigg\{
\bigg({\bf 1}-{{\bf 1}\over g} -{1\over 2}\vec \lambda_n\cdot \vec \lambda_1\bigg)
\bigg({\bf 1}-{{\bf 1}\over g}- {1\over 2}\vec \lambda_1\cdot \vec \lambda_2\bigg)\dots 
\bigg({\bf 1}-{{\bf 1}\over g}- {1\over 2}\vec \lambda_{n-1}\cdot \vec \lambda_n\bigg)
\bigg\} \nonumber \\ &\!\!\!=\!\!\!& {\rm tr}_2\dots  {\rm tr}_n
\bigg\{\bigg[g\bigg({\bf 1}-{{\bf 1}\over g}\bigg)^2 + {1\over 2}\vec \lambda_n\cdot 
\vec \lambda_2\bigg]\dots \bigg({\bf 1}-{{\bf 1}\over g}- {1\over 2}\vec \lambda_{n-1}
\cdot \vec \lambda_n\bigg)\bigg\} \,, \end{eqnarray}
and we have shown the intermediate result of a single flavor-trace, where tr$\,{\bf 1}=g$. 
After taking $n-1$ flavor-traces one arrives at:
\begin{equation}S_n(g) = {\rm tr}_n\bigg\{g^{n-1}\bigg({\bf 1}-{{\bf 1}\over g}\bigg)^n 
+(-1)^n {1\over 2}\vec \lambda_n\cdot \vec \lambda_n \bigg\}\,, \end{equation}
which leads with $\vec \lambda\cdot\vec \lambda=2(g-1/g){\bf 1}$ to the final result: 
\begin{equation}S_n(g) =(g-1)^n +(-1)^n (g^2-1) \,. \end{equation}
One realizes that  $g(g-1)$ is a common divisor of the polynomials $S_n(g)$ of 
degree $n$ in $g$. The 
spin-factors of the 3-, 4- and 5-ring diagrams are: 
\begin{eqnarray}&& S_3(g) =  g(g-1)(g-3)\,, \nonumber \\ && S_4(g) =  g(g-1)(g^2-3g+4)\,,
\nonumber \\ && S_5(g) =  g(g-1)(g^3-4g^2+6g-5) \,. \end{eqnarray}
Although our derivation of the formula for $S_n(g)$ has been elementary, it seems 
to represent a novel result. We note as an aside that the same method applied to 
the symmetrized contact-interaction gives for bosonic $n$-ring diagrams the flavor-factor 
$B_n(g)= (g+1)^2+g^2-1 = (-1)^n S_n(-g)$. 

Let us finally generalize the expressions 
for the resummed interaction energy per particle in eqs.(24,26) to arbitrary 
spin-degeneracy $g$. The sum of the leading $n$-ring diagrams with spin-factors 
$g^n$ takes the form:
\begin{equation}\bar E(k_f)_{\rm lead} = {k_f^2  \alpha^3 \over 4M }  \bigg\{
{g^2 \over 4}\, G\Big({\alpha g \over 2}\Big) \bigg\}\,, \end{equation}
and the complete ring-sum with inclusion of all exchange-type diagrams is 
given by:
\begin{equation}\bar E(k_f) = {k_f^2\alpha^3\over 4M} \bigg\{ {(g-1)^3 \over 4g}  
\, G\Big({\alpha \over 2}(g-1)\Big)- {g^2 -1\over 4g}\, G\Big(-{\alpha \over 2}\Big)
\bigg\} \,, \end{equation}
with the function $G(\alpha)$ defined in eq.(24) and displayed with its 
$\alpha$-dependence in Figs.\,3,4. 
  
\section*{Acknowledgement}
I thank H.W. Hammer, T. Sch\"afer and R. Schmidt for informative discussions.
This work has been supported in part by DFG and NSFC (CRC 110).

\end{document}